# GraphRT: A graph-based deep learning model for predicting the retention time of peptides

**Mark Drvodelic[1], Mingming Gong[1,*], and Andrew I. Webb[2,*]**

1. School of Mathematics and Statistics, The University of Melbourne
2. The Walter and Eliza Hall Institute of Medical Research, Department of Medical Biology, University of Melbourne, Melbourne, Victoria, Australia.
   *Corresponding authors

## Abstract

GraphRT is a graph-based deep learning model that predicts the retention time (RT) of peptides in liquid chromatography tandem mass spectrometry (LC-MS/MS) experiments. Each amino acid is represented as a graph, capturing its atomic and structural properties through a graph neural network. This enables the model to understand not just the chemical composition of each amino acid, but also the intricate relationships between its atoms. The sequential context of the peptide—the order and interaction of amino acids in the sequence is then encoded using recurrent neural networks. This dual approach of graph-based and sequential modelling allows for a comprehensive understanding of both the individual characteristics of amino acids and their collective behaviour in a peptide sequence. GraphRT outperforms all current state-of-the-art models and can predict retention time for peptides containing unseen modifications.

## Main text

Proteomics is the large-scale study of proteins and involves identifying proteins and their chemically modified states within a sample using LC-MS/MS. Peptides, the proteins' digested fragments, are typically separated based on physicochemical properties using reversed-phase chromatography. The amount of time that it takes the molecule to be detected by the MS from the start of the run is known as the retention time (RT) and is contingent on the LC system, elution method and peptide attributes. Recently, RT has been used as a property of peptides to increase the confidence in the matching of MS outputs to peptide sequences. Using predicted RT in conjunction with traditional peptide scoring methods can assist in identifying potential signals and providing higher confidence for a possible match.

Despite the potential utility of RT in aiding peptide identification, current methods for predicting RT values are limited by their inability predict RT for modified forms and for the various LC-MS/MS conditions under which they are analysed. This paper addresses the need for an accurate and robust method to predict peptide RT that can handle a wide range of peptide sequences, including those with post-translational modifications.

The earliest models predicted RT using retention coefficients, an approach that ignored amino acid order and was unsuitable for larger peptides. As an advancement, Petritis et al.[1] leveraged artificial neural networks for peptides exceeding 20 amino acids but lacked sequence-dependent effects consideration.

More recently, DeepRT[2] introduced deep learning to RT prediction, employing convolutional neural networks (CNNs) and long short-term memory (LSTM) networks. This development outperformed its predecessors and accounted for peptide sequence information. Its successor, DeepRT+[3], integrated a capsule network model and transfer learning for handling RT differences across different experimental setups. Both models, however, are limited by their inability to predict RT for peptides with unseen modifications due to their encoding method.

DeepLC, another state-of-the-art model, overcame these limitations, predicting RT for any peptide including those with unseen modifications. It utilised a composite model comprising a CNN for the peptides, a multilayer perceptron (MLP) for global features, and two CNNs for atomic composition. DeepLC's third and fourth sub-models enable accurate RT prediction for unseen modifications by considering the atomic count of amino acids.

Lastly and most recently, AlphaPeptDeep, a Python framework, enabled non-specialists to create models related to proteomics. Its LSTM-based model embeds peptides and atomic counts, predicting the RT of peptides with an arbitrary list of post-translational modifications. The model, when fine-tuned, displayed a slight improvement over DeepLC's performance. Although these models incorporate sequential and atomic information, they disregard structural information about amino acids and peptides.

Here, we introduce GraphRT, a model that leverages a graphical representation of the amino acids that make up a peptide and graph-based deep learning techniques to better capture chemical properties and interactions. This approach outperforms existing models and has the potential to enhance peptide identification accuracy in proteomic workflows. GraphRT is also able to address the limitations of some existing RT prediction methods, particularly their inability to predict RT for modified peptides and varying LC-MS/MS conditions.

Peptides in GraphRT are represented as a sequence of graphs, where each graph corresponds to an amino acid. The encoding process involves first representing each amino acids using its SMILES[4] representation, and then creating a graph object, containing structural and atomic features (such as bonds between atoms and the types and counts of atoms). Computationally, each peptide is represented using three matrices - one encoding atomic features, one encoding edge connectivity, and another encoding edge features. The encoding process is displayed in Figure 1a, for the example amino acids glycine.

The first part of GraphRT, the GNN, processes this sequence-of-graphs representation of the peptide. It converts the sequence of amino acids into a sequence of embedding vectors ($Y_1$, $Y_2$,...,$Y_3$) which provide context for the peptide. This process is shown in Figure 1b. The sequence of embedding vectors is then processed by the second part of GraphRT, an LSTM, which outputs the RT for the sequence of amino acids, the peptide. The overall architecture of GraphRT is shown in Figure 1c.

We suggest that the inclusion of structural information, overlooked by previous models, significantly enhances GraphRT's generalisation ability, leading to superior performance even with unseen modifications. Furthermore, GraphRT's architecture is simpler than models like DeepLC as it does not require secondary information like global peptide properties to achieve excellent results. The work completed here provides a pre-trained version of GraphRT, applicable to any user-provided peptide. However, as RT values are dependent on specific experimental setups, the prediction accuracy is tied to the pre-training dataset. To address this, we calibrate GraphRT's predictions using a calibration dataset corresponding to new experimental setups, ensuring accurate RT prediction. This paper also explores GraphRT's ability to predict RT for unseen modifications, by training it on a peptide set excluding certain modifications and testing it on a peptide set containing only those modifications.

This work makes several key contributions. First, GraphRT is introducted, a novel model that surpasses accuracy of all prior models in predicting peptide RT across diverse datasets and LC-MS/MS conditions. Second, we demonstrate GraphRT's unique ability to generalise to unseen modifications, thereby enhancing the accuracy of peptide identification in proteomic workflows. Finally, we present a Python package that enables researchers and biologists to easily integrate GraphRT into their proteomic analyses, making this innovative model accessible and beneficial to the broader scientific community.

GraphRT was trained and tested on 20 different datasets, shown in Supplementary Table 2, replicating the comparisons described in the DeepLC[5] paper. The results were then compared with DeepLC, AlphaPeptDeep, and where possible, DeepRT+. GraphRT outperforms all aforementioned models, DeepLC, AlphaPeptDeep, and DeepRT+, using its graphical encoding of peptides. Figure 2 illustrates this, showing GraphRT performing better than DeepLC and AlphaPeptDeep, especially in datasets with larger mean absolute errors (MAEs), such as Yeast 1h/2h, Pancreas, HeLa DeepRT, Arabidopsis, HeLa Lumos 1h/2h, Plasma Lumos 2h and ProteomeTools/PTM. For datasets with lower MAEs, the difference in performance is much smaller. AlphaPeptDeep seems to struggle with larger datasets like SWATH library, ProteomeTools, and DIAHF, suggesting potential limitations in its handling of larger datasets.

GraphRT also outperforms DeepRT+ in terms of the $\Delta t_{95\%}$ metric across 8 of the datasets compared, as seen in Supplementary Table 2. GraphRT also exhibits the highest r values for five of these datasets, implying better fit. In the case of HeLa DeepRT, despite GraphRT's slightly lower r, its superior $\Delta t_{95\%}$ suggests fewer outliers and closer results to the line of perfect fit compared to DeepRT+. Supplementary Table 2 provides the exact MAE, r and $\delta t_{95\%}$ metrics for each dataset, for each model. Additionally, the graphs shown in Supplementary Figures 1 and 2 demonstrate GraphRT's performance plotting true versus predicted of MAE values for each of the datasets.

GraphRT's ability to generalise to unseen modifications was evaluated using a subset of the ProteomeTools PTM (PTPTM) dataset. This dataset comprises approximately 85% peptides with one of 14 unique modifications and was split into 14 corresponding data subsets to analyse each modification separately. These subsets are shown in Supplementary Table 3. Each subset was divided into training, validation, and holdout test sets, the latter containing only peptides with a specific modification. An additional "unencoded" holdout test set was created, which did not encode the peptide modifications. This was done to assess if encoding modifications into input peptides improves RT prediction accuracy.

Subsequently, different versions of GraphRT, DeepLC, and AlphaPeptDeep were trained using corresponding training sets. Their performance was evaluated on both encoded and unencoded test sets evaluating performance in predicting RT for a particular modification. The graphs shown in Supplementary Figures 3, 4, 5 and 6 show GraphRT's true versus predicted plots of MAE values for each of the modification datasets. Supplementary Figure 7 and Supplementary Table 4 demonstrate GraphRT successfully generalised for most of the modification datasets, frequently surpassing the performance of other models significantly. GraphRT outperformed the other two models in terms of MAE for 11 out of 14 modification datasets - all those except methyl, phospho and trimethyl. For 10 of these 11 modifications, including dimethyl, acetyl, succinyl, propionyl, crotonyl, malonyl, formyl, oxidation, carbamidomethyl, and deamidated, GraphRT performed better when modifications were encoded rather than ignored.

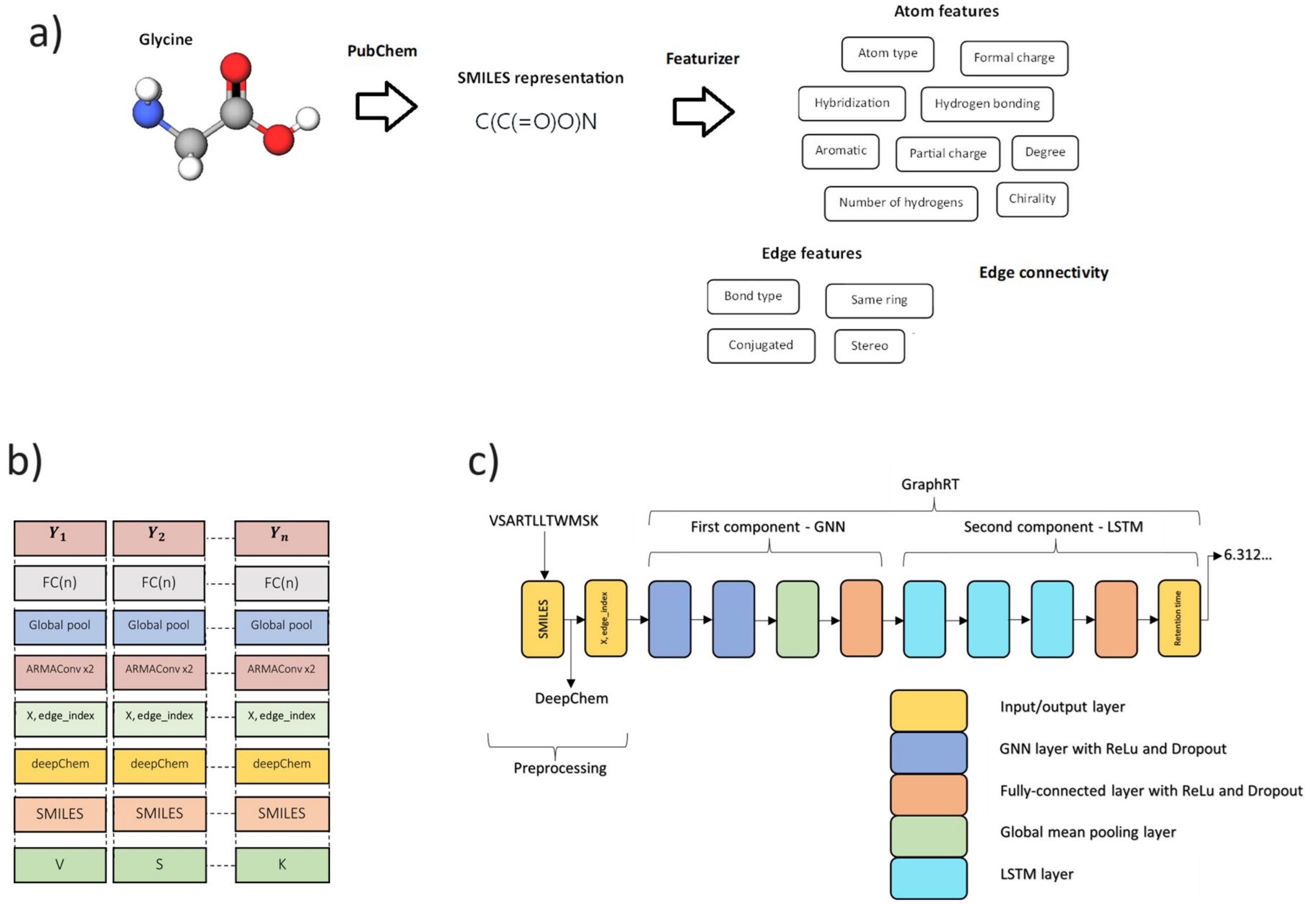

**Figure 1.** a) The encoding process by which a single amino acid (here, glycine) is converted into a set of feature matrices by using the SMILES representation as a proxy set. b) An input peptide, represented by a sequence of amino acids, is first encoded into feature matrices, and then each amino acid is individually processed by a GNN to create a sequence of embedding vectors. c) The overall structure of GraphRT, which is divided into two main sections, the GNN, which embeds each amino acid in the peptide separately, and the LSTM, which processes the entire peptide to produce the RT output.

The study revealed that GraphRT not only learns the additive contribution of a modified amino acid to the overall RT but also its interactive contribution, based on the amino acid's position in the peptide. This learning was less apparent in DeepLC and AlphaPeptDeep. Certain modifications, like methyl, oxidation, carbamidomethyl, and deamidated, showed less significant improvement with encoding, suggesting these modifications benefit less from encoding. Interestingly, GraphRT performance dropped for phospho and nitro when modifications were encoded, indicating these modifications might be too chemically distinct from others to benefit from encoding. AlphaPeptDeep's unencoded results surpassed encoded ones for seven modifications, hinting at less effective encoding techniques in AlphaPeptDeep compared to GraphRT and DeepLC. However, for trimethyl, DeepLC and AlphaPeptDeep significantly outperformed GraphRT, suggesting possible encoding issues for trimethyl in GraphRT. Furthermore, trimethyl's unencoded test set performed better, indicating potential negative effects of encoding for this modification.

We also evaluated the feasibility of a given GraphRT instance to accurately predict Retention Time (RT) for peptides from different Liquid Chromatography (LC) setups. Results showed that when the LC column type of the pre-trained model and test set were identical, there was a high positive correlation in RT order despite poor performance in terms of MAE. This is displayed in Supplementary Figures 8 and 9, which show the MAE and correlation r for each version of GraphRT tested on the hold out test sets of the other 19 data sets. The x-axis shows the data set GraphRT is pre-trained on, and the y-axis shows one of the other 19 data sets from which the holdout test set originates. This result implies that the peptide's RT order is maintained when using the same column type, though the absolute value may vary significantly.

These results indicate that direct generalisation to different datasets is unachievable without calibration. However, if the pre-trained GraphRT model and queried peptide dataset use the same column type, transformation could significantly decrease the MAE. When we tested a simple linear transformation using two points sampled from the dataset, applying the transformation, the MAE was significantly reduced, confirming the effectiveness of the proposed calibration method.

The power of GraphRT resides in its methodology. By first utilising the SMILES representation of an amino acid, and then converting them into graphs, GraphRT can effectively capture structural information of each amino acid beyond what traditional chemical formulas can provide. By then representing each peptide as a sequence of these graphs, GraphRT also captures crucial sequential information about a peptide which is vital to determining its RT. GraphRT presents a breakthrough in the prediction of peptide RT by utilising a graph-based deep learning approach and GNN-LSTM architecture to provide an efficient and more accurate tool for researchers in proteomics. Its proficiency in dealing with unseen modifications and adaptation to different datasets using a minimal set of calibration peptides also eliminates the need for a multitude of models, paving the way for more efficient and streamlined proteomics research.

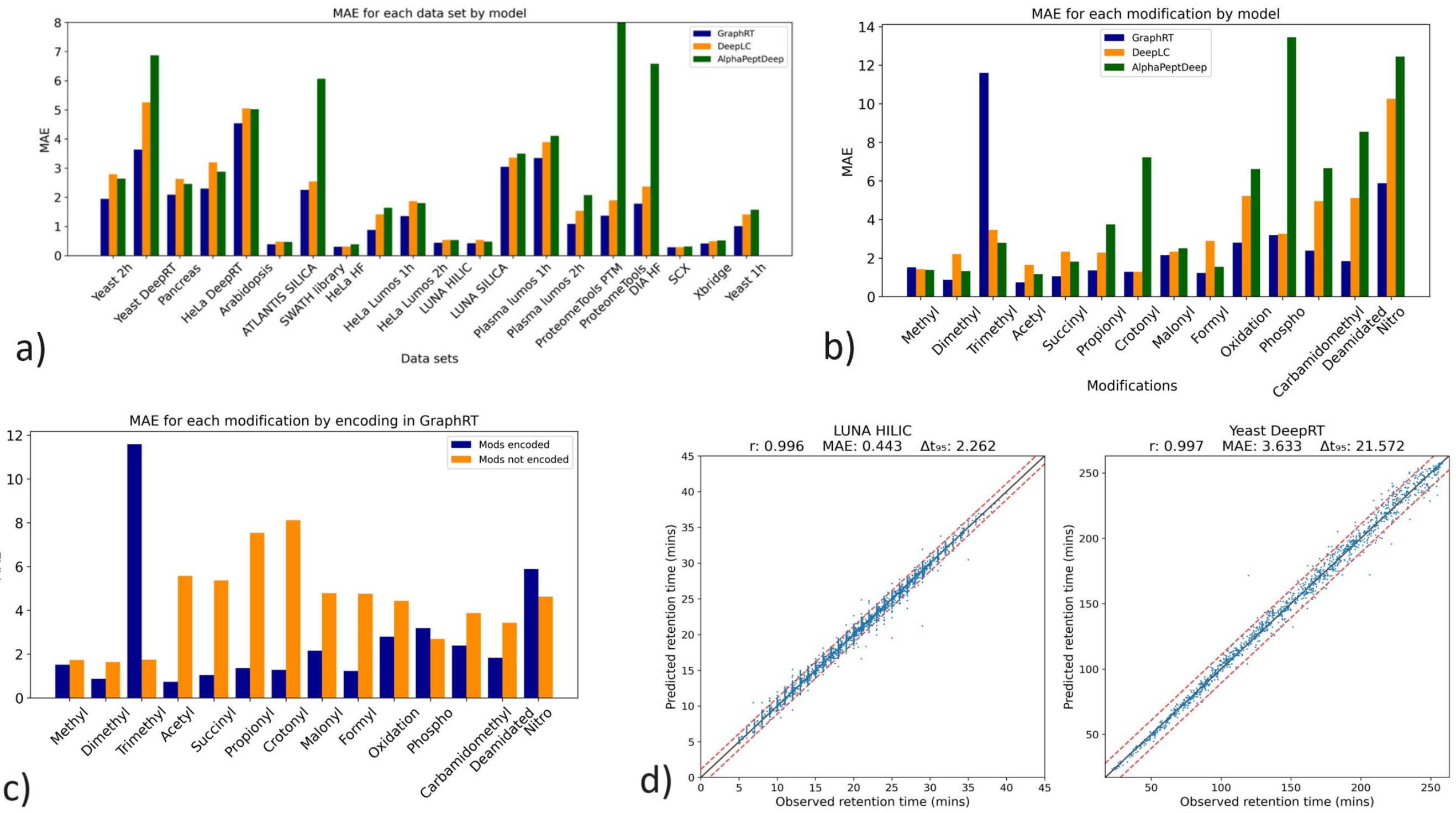

**Figure 2.** a) Bar chart of the MAE achieved by GraphRT, DeepLC and AlphaPeptDeep, for all 20 data sets. GraphRT performs better than both DeepLC and AlphaPeptDeep for 19 of the 20 data sets. b) Bar chart of the MAE achieved by GraphRT, DeepLC and AlphaPeptDeep when modifications are encoded, for all 14 modification data sets. GraphRT outperforms DeepLC and AlphaPeptDeep for 11 of the 14 modifications. c) Bar chart of the MAE for all 14 modifications when modifications are encoded or not encoded in the amino acids. In 11 of the 14 modifications, encoding modifications improves the result. d) Example of the predicted versus true plot for the Yeast DeepRT data set and LUNA HILIC data set.

## Methods

### Peptide encoding

Most recent models do not incorporate structural information about an atom and are based solely on the sequential information of amino acids in a peptide. These models are varied in their architecture, but most utilise a one-hot encoding of the peptides as input. DeepLC[5] found a way to incorporate not just sequential information but the atomic composition of amino acids as input into their model. However, structural information about the peptide is disregarded. GraphRT represents amino acids as graphs, and peptides as a sequence of amino acid graphs. This is a natural representation given that an amino acid is a molecule, and molecules can easily be represented as graphs. By representing amino acids as graphs, we can incorporate structural information about the amino acid, such as the bonds between atoms in the amino acid and atomic composition, such as the types of atoms that make up the amino acid.

The first step to converting an amino acid into a graphical representation involves utilising its SMILES representation. Every single molecule can be represented by an ASCII string, known as its SMILES[4]. SMILES is a chemical notation system that represents chemical structure using a text-based format. It is a standard way of representing molecular structures so that they are computer readable. SMILES representations were created using a set of rules to deconstruct the graphical representation of a molecule, such as the fact that double bonds are represented by a '='. As a result, unlike chemical formulas which only provide information about the types and number of atoms in a molecule, SMILES can provide structural information. For example, the chemical formula $C_2H_4O_3$ describes a molecule containing 2 carbon atoms, 4 hydrogen atoms, and 3 oxygen atoms. This representation does not tell us how these atoms are connected or arranged in space, however. In contrast, a SMILES string representation with the same chemical formula, "C(O)C(=O)O", encodes the connectivity and stereochemistry of a molecule, giving a more detailed representation of its structure. For these reasons, converting between structural representations and SMILES is simple. SMILES representations were indeed created using a set of rules to deconstruct the graphical representation of a molecule, such as the fact that double bonds are represented by a '='.

For each amino acid (modified or unmodified) we find a SMILES representation using the online database of chemical molecules, PubChem[6]. Table 1 contains all the SMILES representations for all the amino acids that exist in the data sets analysed here. Once every amino acid has been matched to a corresponding SMILES representation, the amino acids can be converted into a graph. For any given peptide, we convert each of its constituent amino acids to graphs, so that GraphRT, and the GNN component of GraphRT, can process it. To simplify this conversion, we employ DeepChem[7], a Python library focused on applying deep learning to chemistry. We utilise DeepChem's "MolGraphConvFeatuizer" (MGCF) which is a function that automates the conversion of the SMILES of a molecule into its graphical representation. It does this by algorithmically enacting all the rules originally stipulated in the original SMILES paper[4]. For more details the reader is encouraged to explore the original SMILES paper[4] and the DeepChem paper[7].

The graphical output of MGCF is a set of nodes, which are the atoms in the amino acid, and edges, which are the chemical bonds between atoms in the amino acid[8]. Computationally in Python it is a "GraphData" object. This object stores features of the graphical representation of the molecule. The two features we utilise here are the node feature matrix and the edge feature matrix. The node feature matrix, $X \in R^{Nnodes \times Nfeatures}$, stores the feature vectors for each node (that is, each atom) and has the dimensions number of nodes $N_{nodes}$, by number of features $N_{features}$. In GraphRT, $N_{nodes}$ depends on the amino acid and different amino acids will have a different number of atoms. The number of features for each node, $N_{features}$, is 31. These features include: a one-hot vector to represent the chemical element of the atom (out of the elements carbon, nitrogen, oxygen, fluorine, phosphorus, sulfur, chlorine, bromine, iodine, and all other atoms); formal charge (the charge assigned to an atom in a molecule); hybridization[9]; hydrogen bonding (a one-hot vector of whether this atom is a hydrogen bond donor or acceptor); aromatic (a one-hot vector of whether the atom belongs to an aromatic ring); degree (a one-hot vector giving the degree of the atom); number of hydrogens (a one-hot vector giving the number of hydrogens that the atom is connected to); and the partial charge (is the charge of an atom caused by asymmetric distribution of electrons in chemical bonds). The edge index matrix edge_index $\in R^{2\times N_{edges}}$, stores the graph connectivity information and has dimensions 2 by the number of edges, $N_{edges}$, in the molecule. An example of these two feature matrices is shown below in matrix format:

$$X = \begin{bmatrix} 5 & 1 & \cdots \\ \vdots & \ddots & \\ 0.4 & & 3 \end{bmatrix}; \qquad edge_index = \begin{bmatrix} 2 & \cdots & 4 \\ 1 & \cdots & 7 \end{bmatrix}$$

**Architecture**

In the previous section, a method was described to encode peptides so that they can be easily processed by a deep learning model. Peptides, represented as a string of characters corresponding to each amino acid, were converted into a sequence of graphs. This sequence of graphs representation can be readily inputted into GraphRT. Here, the architecture of GraphRT is described.

GraphRT is structured as a GNN-LSTM hybrid model. The inputs to GraphRT are a sequence of “GraphData" objects, being composed of a node feature matrix X and an edge index matrix edge_index. Each graph object in the sequence is inputted into the first component of GraphRT - a GNN with two ARMAConv layers - one at a time. X is modified as it passes through each layer of the GNN and edge_index is used to weight those updates accordingly. The output of these layers is a new feature matrix, $\bar{X}$ , representing new features for each node in the input graph, updated according to their neighbouring nodes. To generate a final feature vector for each amino acid making up the peptide (which can then be processed by the rest of GraphRT, rather than the entire matrix $\bar{X}$ ), this output layer of the GNN is then processed by a global mean pooling layer. The global mean pooling layer averages over all nodes in the matrix to produce a single feature vector for each input graph in the peptide sequence. This feature vector is then connected to an MLP which outputs another final vector - the embedding vector. The result of this first component of GraphRT is a sequence of embedding vectors, representing embeddings for each graph in the original input sequence. That is, each embedding vector summarises a particular amino acid in the original peptide. This sequence of embedding vectors can then be processed by the second part of GraphRT - the LSTM. Figure 1b summarises the GNN component of GraphRT for the example peptide VSARTLLTWMSK, as example input.

After the GNN component of GraphRT, the sequence of graphs representing the input peptide has been converted into a sequence of embedding vectors ($Y_1,...,Y_n$). This representation is then inputted into the second part of GraphRT - the LSTM. A bidirectional LSTM with n = 4 layers is used. The final layer of the LSTM is extracted, connected to a fully connected MLP, and then a final output layer outputs a single regression value - the RT. Figure 5 provides of overview of the network architecture.

The GraphRT models here were created and trained using PyTorch[10] and Python[11]. The hyperparameters were chosen using a combination of prior knowledge and hyperparameter tuning over a set of specified values. For each training epoch the model was exposed to the entire training set via a set of randomly generated, same-length peptide batches, with the batches shuffled every epoch. The average training loss and validation loss were calculated at the end of every epoch. The training loss was added incrementally during training, and then averaged after training. In contrast, the validation loss was calculated post-training using the validation batches, every epoch.

The total number of epochs was 300, however, the actual number of epoch experienced was generally less than this because of the early stopping hyperparameter, n. If there was no improvement in the average validation loss for n epochs, then training would cease. The final model selected was that which minimized the average validation loss over all training epochs. This model was then used to predict the RT for all peptides in the holdout test set.

The results for DeepLC were obtained directly from the paper. On the other hand, the results for AlphaPeptDeep were obtained by training and testing the 'AlphaRTModel' model class, an untrained, base model used by AlphaPeptDeep for RT predition, on each data set described in Table 2. All hyperparameters were kept the same as those used in the AlphaPeptDeep paper, except for the number of training epochs which was increased as this produced better results. These hyperparameter values used were epoch=300, warmup epoch=30, lr=1e–4, dropout=0.1, mini-batch size=256.

Three metrics were used to measure performance and compare the models. Firstly, the mean absolute error, or MAE, which is the arithmetic mean of the absolute difference between true and predicted RT. MAE is used to gauge how accurate the actual predictions are and how close the predicted RT values are to the observed values. Next is the Pearson correlation coefficient, r, between the true and predicted RT values. The correlation coefficient is useful as it describes how close to a line the data lies. When r = ±1, the data forms a perfect line. The line formed when r = ±1, however, is not necessarily the line of perfect prediction. Such a line is given by $\hat{y} = y$, where predicted ($\hat{y}$) and observed (y) RT are identical, and for which MAE = 0.

For example, we might observe a very high r value (close to 1) and a very large MAE value (very inaccurate results). This outcome might indicate that predictions are "relatively" correct but offset by some linear amount and in need of recalibration. In this sense, taken together MAE and r complement one another and show how close the results are to a perfect fit. The formulas for both MAE and r respectively are given by

$$\text{MAE} = \frac{1}{n}\sum_{i=1}^{n}|\hat{y}_i - y_i| \quad \text{and} \tag{1}$$

$$r = \frac{\sum_{i=1}^{n}(y_i - \bar{y})(\hat{y}_i - \bar{\hat{y}})}{\sqrt{\sum_{i=1}^{n}(y_i - \bar{y})^2(\hat{y}_i - \bar{\hat{y}})^2}}, \tag{2}$$

where $\hat{y}_i$ and $y_i$ are the predicted and true RT values for the $i^{th}$ data point, respectively. Also, $\bar{y}$ and $\bar{\hat{y}}$ are the corresponding means of the predicted and true RT values, respectively.

The final evaluation metric was the $\Delta t_{95\%}$ metric. It is two times the 0.95 quantile of the absolute RT error, $|\hat{y}_i - y_i|$ and can be thought of as a symmetric error bound that contains 95% of the error distribution. It is used to gauge the spread of the errors similarly to r, but it is more robust to extreme outliers than r. For example, suppose that the very extreme outliers (those in the top 5%) were shifted further away from $\hat{y} = y$. This would result in only a very slight change in $\Delta t_{95\%}$, but a very large change in r.

| Amino acid | SMILES |
|---|---|
| Glycine (G) | NCC(=O)O |
| Alanine (A) | N[C@@]([H])(C)C(=O)O |
| Arginine (R) | N[C@@]([H])(CCCNC(=N)N)C(=O)O |
| Asparagine (N) | N[C@@]([H])(CC(=O)N)C(=O)O |
| Aspartic acid (D) | N[C@@]([H])(CC(=O)O)C(=O)O |
| Cysteine (C) | N[C@@]([H])(CS)C(=O)O |
| Glutamic acid (E) | N[C@@]([H])(CCC(=O)O)C(=O)O |
| Glutamine (Q) | N[C@@]([H])(CCC(=O)N)C(=O)O |
| Histidine (H) | N[C@@]([H])(CC1=CN=C-N1)C(=O)O |
| Isoleucine (I) | N[C@@]([H])(C(CC)C)C(=O)O |
| Leucine (L) | N[C@@]([H])(CC(C)C)C(=O)O |
| Lysine (K) | N[C@@]([H])(CCCCN)C(=O)O |
| Methionine (M) | N[C@@]([H])(CCSC)C(=O)O |
| Phenylalanine (F) | N[C@@]([H])(Cc1ccccc1)C(=O)O |
| Proline (P) | N1[C@@]([H])(CCC1)C(=O)O |
| Serine (S) | N[C@@]([H])(CO)C(=O)O |
| Threonine (T) | N[C@@]([H])(C(O)C)C(=O)O |
| Tryptophan (W) | N[C@@]([H])(CC(=CN2)C1=C2C=CC=C1)C(=O)O |
| Tyrosine (Y) | N[C@@]([H])(Cc1ccc(O)cc1)C(=O)O |
| Valine (V) | N[C@@]([H])(C(C)C)C(=O)O |
| R Methyl | CN=C(N)NCCCC(C(=O)O)N |
| R Dimethyl | CN(C)C(CCCN=C(N)N)C(=O)O |
| C Carbamidomethyl | C(C(C(=O)O)N)SCC(=O)N |
| K Methyl | CNCCCCC(C(=O)O)N |
| K Dimethyl | CN(C)CCCCC(C(=O)O)N |
| K Trimethyl | C[N+](C)(C)CCCCC(C(=O)O)N |
| K Crotonyl | CC=CC(=O)NCCCCC(C(=O)O)N |
| K Succinyl | C(CCNC(=O)CCC(=O)[O-])CC(C(=O)O)N |
| K Propionyl | CCC(=O)NCCCCC(C(=O)O)N |
| K Formyl | C(CCNC=O)CC(C(=O)O)N |
| M Oxidation | CS(=O)CCC(C(=O)O)N |
| S Phospho | C(C(C(=O)O)N)OP(=O)(O)O |
| T Phospho | CC(C(C(=O)O)NP(=O)(O)O)O |
| Y Phospho | C1=CC(=CC=C1CC(C(=O)O)N)OP(=O)(O)O |
| N Acetyl | C[C]=O |
| K Acetyl | CC(=O)NCCCCC(C(=O)O)N |
| Y Nitro | C1=CC(=C(C=C1CC(C(=O)O)N)[N+](=O)[O-])O |
| P Oxidation | C1CC(=NC1)C(=O)O |
| R Deamidated | C(CC(C(=O)O)N)CNC(=O)N |
| K Malonyl | C(*)([C@@H](N*)CCCCNC(CC([O-])=O)=O)=O |

**Table 1.** The amino acid and corresponding SMILES representation.

## Supplementary Information

| Data sets | # train | # val | # test | total |
|---|---|---|---|---|
| Yeast 2h | 23512 | 1237 | 2750 | 27499 |
| Yeast DeepRT | 12197 | 641 | 1426 | 14264 |
| Pancreas | 43002 | 2263 | 5029 | 50294 |
| HeLa DeepRT | 2917 | 153 | 341 | 3411 |
| Arabidopsis | 13310 | 700 | 1556 | 15566 |
| ATLANTIS SILICA | 33421 | 1759 | 3909 | 39089 |
| SWATH library | 43080 | 2239 | 5049 | 50368 |
| HeLa HF | 137821 | 7253 | 16119 | 161193 |
| HeLa Lumos 1h | 21638 | 1138 | 2530 | 25306 |
| HeLa Lumos 2h | 49495 | 2604 | 5788 | 57887 |
| LUNA HILIC | 30848 | 1623 | 3607 | 36078 |
| LUNA SILICA | 31483 | 1656 | 3682 | 36821 |
| Plasma lumos 1h | 2997 | 157 | 350 | 3504 |
| Plasma lumos 2h | 3659 | 192 | 428 | 4279 |
| ProteomeTools PTM | 4867 | 256 | 569 | 5692 |
| ProteomeTools | 133482 | 7025 | 15611 | 156118 |
| DIA HF | 43227 | 2413 | 4726 | 50366 |
| SCX | 26051 | 1371 | 3047 | 30469 |
| Xbridge | 34231 | 1801 | 4003 | 40035 |
| Yeast 1h | 15822 | 832 | 1850 | 18504 |

**Supplementary Table 1.** The number of training, validation, testing and total peptides in each of the data sets.

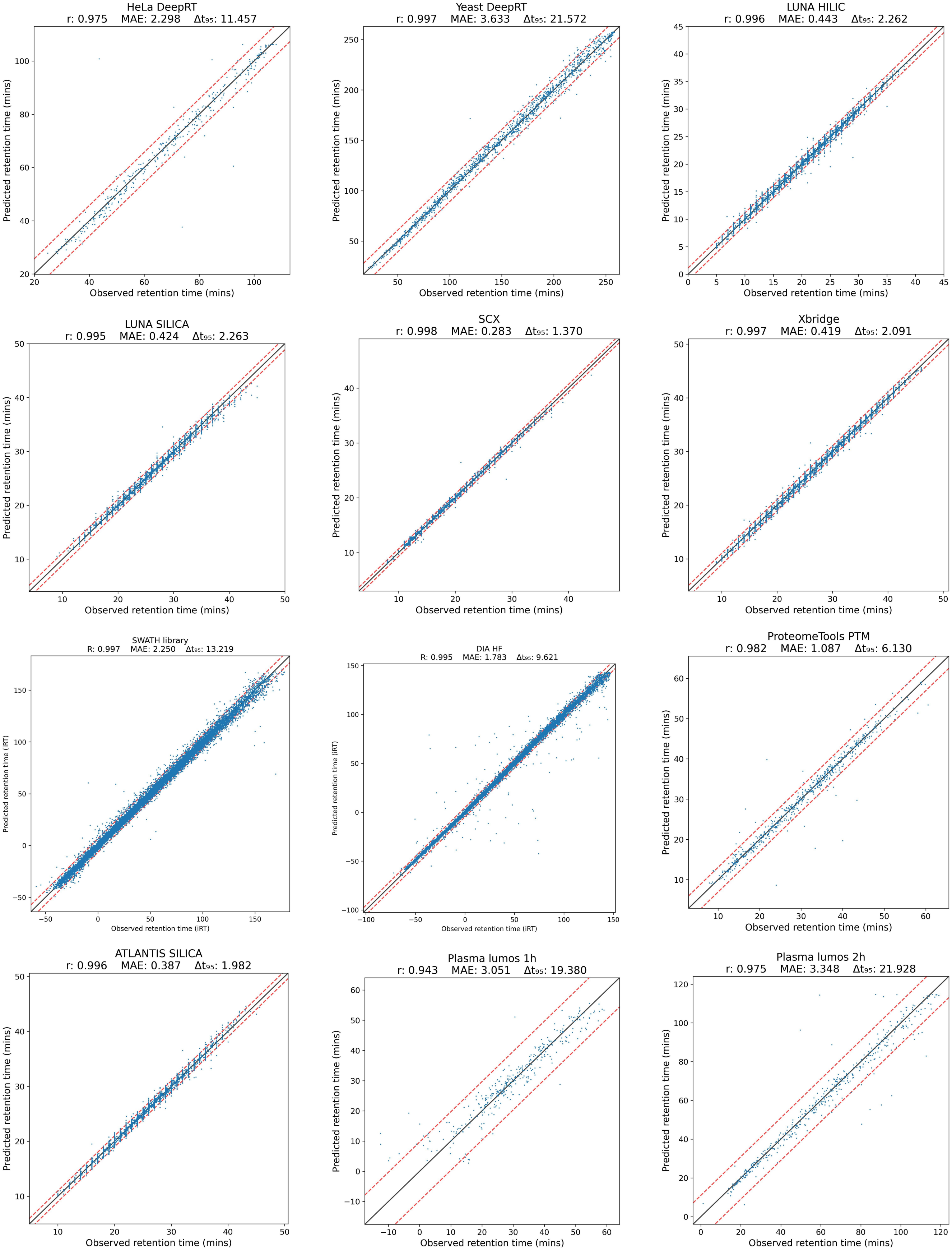

**Supplementary Figure 1.** Plots of true versus predicted RT for GraphRT.

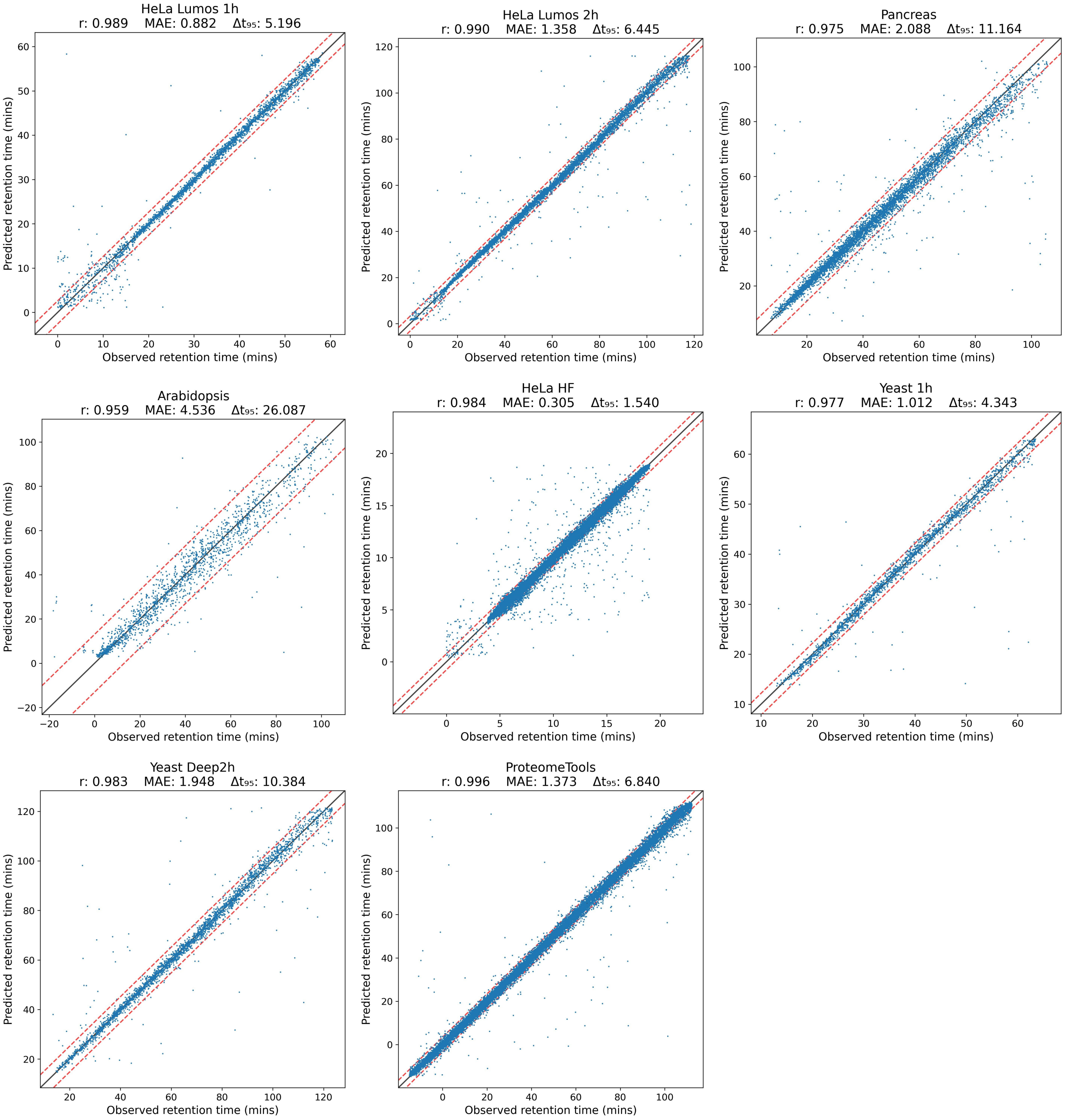

**Supplementary Figure 2.** Plots of true versus predicted RT for GraphRT.

| Models | GraphRT | | | DeepLC | | | AlphaPeptDeep | | | DeepRT+ | | |
|---|---|---|---|---|---|---|---|---|---|---|---|---|
| Data sets | MAE | *r* | $\Delta t_{95\%}$ | MAE | *r* | $\Delta t_{95\%}$ | MAE | *r* | $\Delta t_{95\%}$ | MAE | *r* | $\Delta t_{95\%}$ |
| Yeast 2h | 1.95 | 0.983 | 10.38 | 2.79 | 0.975 | 17.34 | 2.64 | 0.980 | 14.01 | | | |
| Yeast DeepRT | 3.63 | 0.997 | 21.57 | 5.26 | 0.993 | 32.18 | 6.87 | 0.988 | 39.66 | | 0.993 | 25.88 |
| Pancreas | 2.09 | 0.975 | 11.16 | 2.63 | 0.967 | 14.22 | 2.46 | 0.971 | 13.05 | | | |
| HeLa DeepRT | 2.30 | 0.975 | 11.46 | 3.20 | 0.967 | 16.73 | 2.88 | 0.971 | 15.50 | | 0.985 | 12.56 |
| SCX | 0.28 | 0.998 | 1.37 | 0.29 | 0.998 | 1.57 | 0.31 | 0.992 | 1.45 | | 0.998 | 1.42 |
| Arabidopsis | 4.54 | 0.959 | 26.09 | 5.05 | 0.951 | 27.74 | 5.02 | 0.953 | 28.68 | | | |
| Atlantis Silica | 0.39 | 0.996 | 1.98 | 0.48 | 0.993 | 2.56 | 0.47 | 0.993 | 2.40 | | 0.995 | 2.10 |
| SWATH library | 2.25 | 0.997 | 13.22 | 2.54 | 0.997 | 14.88 | 6.07 | 0.984 | 49.20 | | 0.997 | 13.40 |
| HeLa HF | 0.31 | 0.984 | 1.54 | 0.31 | 0.984 | 1.62 | 0.39 | 0.980 | 1.83 | | | |
| HeLa Lumos 1h | 0.88 | 0.989 | 5.20 | 1.41 | 0.983 | 10.39 | 1.64 | 0.985 | 9.63 | | | |
| HeLa Lumos 2h | 1.36 | 0.990 | 6.45 | 1.87 | 0.986 | 10.33 | 1.80 | 0.987 | 9.50 | | | |
| Xbridge | 0.42 | 0.997 | 2.09 | 0.49 | 0.996 | 2.64 | 0.52 | 0.996 | 2.57 | | 0.996 | 2.36 |
| Yeast 1h | 1.01 | 0.977 | 4.34 | 1.41 | 0.971 | 8.19 | 1.57 | 0.971 | 8.01 | | | |
| DIA HF | 1.78 | 0.995 | 9.62 | 2.37 | 0.994 | 13.28 | 6.58 | 0.969 | 67.58 | | | |
| ProteomeTools | 1.37 | 0.996 | 6.84 | 1.9 | 0.995 | 9.96 | 23.82 | 0.996 | 117.73 | | | |
| ProteomeTools PTM | 1.09 | 0.982 | 6.13 | 1.54 | 0.975 | 8.51 | 2.08 | 0.961 | 11.03 | | | |
| Plasma Lumos 1h | 3.05 | 0.943 | 19.38 | 3.36 | 0.938 | 19.22 | 3.50 | 0.942 | 20.63 | | | |
| Plasma Lumos 2h | 3.35 | 0.975 | 21.93 | 3.89 | 0.968 | 20.51 | 4.11 | 0.970 | 23.77 | | | |
| Luna Hilic | 0.44 | 0.996 | 2.26 | 0.54 | 0.994 | 2.82 | 0.53 | 0.994 | 2.70 | | 0.994 | 2.55 |
| Luna Silica | 0.42 | 0.995 | 2.26 | 0.54 | 0.991 | 2.99 | 0.48 | 0.993 | 2.55 | | 0.994 | 2.30 |

**Supplementary Table 2.** Comparison of GraphRT, DeepLC, AlphaPeptDeep and DeepRT+ for each of the 20 data sets.

| # samples in each set | | | | |
|---|---|---|---|---|
| Modification | Amino acids | Training | Validation | Test |
| Methyl | R,K | 4696 | 247 | 751 |
| Dimethyl | R,K | 4771 | 251 | 672 |
| Trimethyl | K | 5083 | 267 | 344 |
| Acetyl | N,K | 5183 | 272 | 239 |
| Succinyl | K | 5091 | 267 | 336 |
| Propionyl | K | 5085 | 267 | 342 |
| Crotonyl | K | 5112 | 269 | 313 |
| Malonyl | K | 5112 | 269 | 313 |
| Formyl | K | 5051 | 265 | 378 |
| Oxidation | M,P | 4277 | 225 | 1192 |
| Phospho | S,T,Y | 5210 | 274 | 210 |
| Carbamidomethyl | C | 4969 | 261 | 464 |
| Deamidated | R | 5200 | 273 | 221 |
| Nitro | Y | 5209 | 274 | 211 |

**Supplementary Table 3.** Each of the 14 modifications analysed in this section, with the corresponding amino acids they act on, as well as the number of samples in each of the training, validation and test sets, generated from the PTPTM data set.

| Model | GraphRT | | | | DeepLC | | | | AlphaPeptDeep | | | |
|---|---|---|---|---|---|---|---|---|---|---|---|---|
| Mod encoding | Encoded | | Unencoded | | Encoded | | Unencoded | | Encoded | | Unencoded | |
| Modification | MAE | *r* | MAE | *r* | MAE | *r* | MAE | *r* | MAE | *r* | MAE | *r* |
| Methyl | 1.52 | 0.974 | 1.74 | 0.952 | 1.43 | 0.974 | 1.79 | 0.967 | 1.38 | 0.969 | 3.43 | 0.946 |
| Dimethyl | 0.88 | 0.993 | 1.64 | 0.965 | 2.20 | 0.964 | 1.52 | 0.979 | 1.32 | 0.981 | 3.13 | 0.956 |
| Trimethyl | 11.60 | 0.908 | 1.75 | 0.951 | 3.47 | 0.968 | 0.95 | 0.982 | 2.79 | 0.965 | 4.61 | 0.947 |
| Acetyl | 0.74 | 0.995 | 5.58 | 0.954 | 1.64 | 0.981 | 5.94 | 0.982 | 1.17 | 0.989 | 2.42 | 0.973 |
| Succinyl | 1.05 | 0.995 | 5.37 | 0.965 | 2.33 | 0.974 | 6.01 | 0.985 | 1.81 | 0.980 | 3.15 | 0.976 |
| Propionyl | 1.36 | 0.996 | 7.54 | 0.960 | 2.28 | 0.988 | 7.56 | 0.973 | 3.74 | 0.987 | 4.93 | 0.968 |
| Crotonyl | 1.286 | 0.988 | 8.13 | 0.952 | 1.288 | 0.990 | 9.26 | 0.974 | 7.22 | 0.966 | 6.21 | 0.961 |
| Malonyl | 2.16 | 0.959 | 4.79 | 0.928 | 2.33 | 0.951 | 5.47 | 0.953 | 2.51 | 0.943 | 2.62 | 0.927 |
| Formyl | 1.23 | 0.976 | 4.75 | 0.952 | 2.89 | 0.958 | 4.95 | 0.964 | 1.55 | 0.970 | 2.61 | 0.964 |
| Oxidation | 2.80 | 0.949 | 4.43 | 0.886 | 5.22 | 0.917 | 6.19 | 0.875 | 6.62 | 0.919 | 5.99 | 0.893 |
| Phospho | 3.19 | 0.967 | 2.69 | 0.963 | 3.26 | 0.962 | 3.59 | 0.955 | 13.44 | 0.953 | 2.64 | 0.968 |
| Carbamidomethyl | 2.39 | 0.923 | 3.88 | 0.877 | 4.95 | 0.822 | 5.01 | 0.810 | 6.66 | 0.879 | 3.88 | 0.869 |
| Deamidated | 1.83 | 0.990 | 3.43 | 0.967 | 5.11 | 0.970 | 3.72 | 0.975 | 8.54 | 0.946 | 3.60 | 0.983 |
| Nitro | 5.88 | 0.949 | 4.63 | 0.939 | 10.25 | 0.926 | 5.78 | 0.952 | 12.44 | 0.932 | 5.31 | 0.949 |

**Supplementary Table 4.** Comparison of the MAE and r value between GraphRT, DeepLC and AlphaPeptDeep in the case of modifications being encoded, or not encoded, in the holdout test set. We see that in 11 out of 14 of the modifications, GraphRT performs better than DeepLC. The best MAE for each row (modification) is boldfaced.

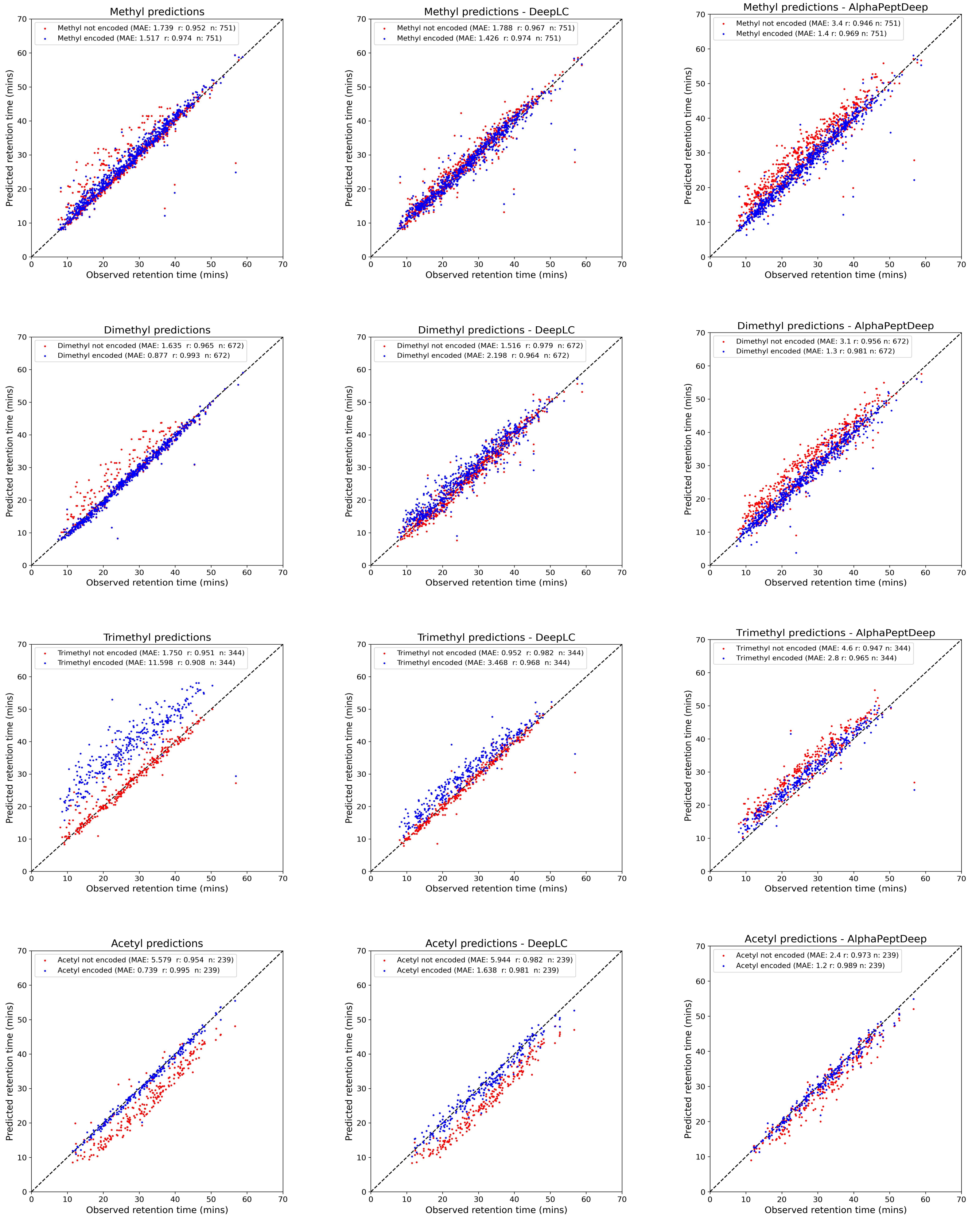

**Supplementary Figure 3.** Plot of predicted vs. true RT for the modifications methyl, dimethyl, trimethyl and acetyl in each row, using GraphRT, DeepLC and AlphaPeptDeep in each column.

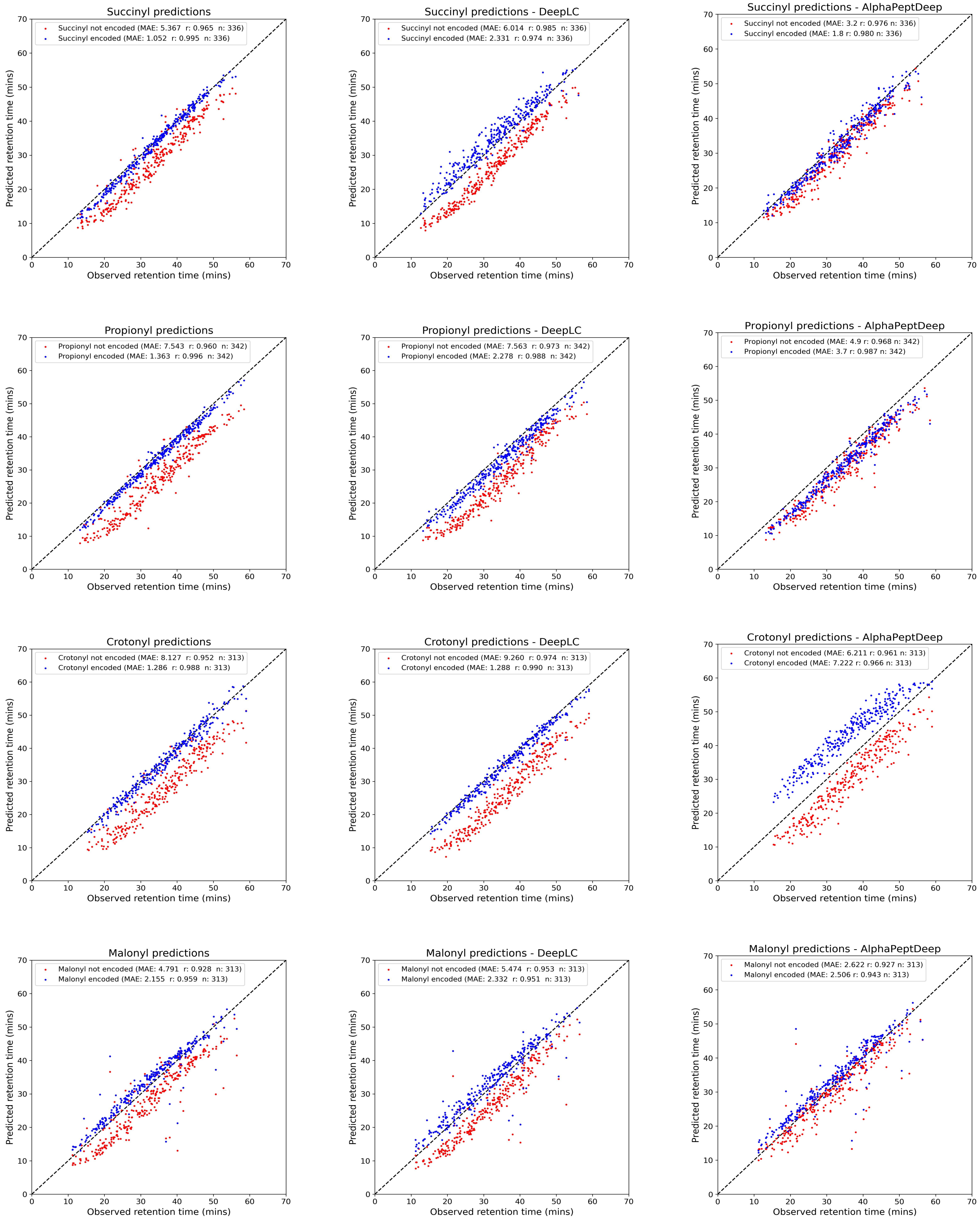

**Supplementary Figure 4.** Plot of predicted vs. true RT for the modifications succinyl, propionyl, crotonyl and malonyl in each row, using GraphRT, DeepLC and AlphaPeptDeep in each column.

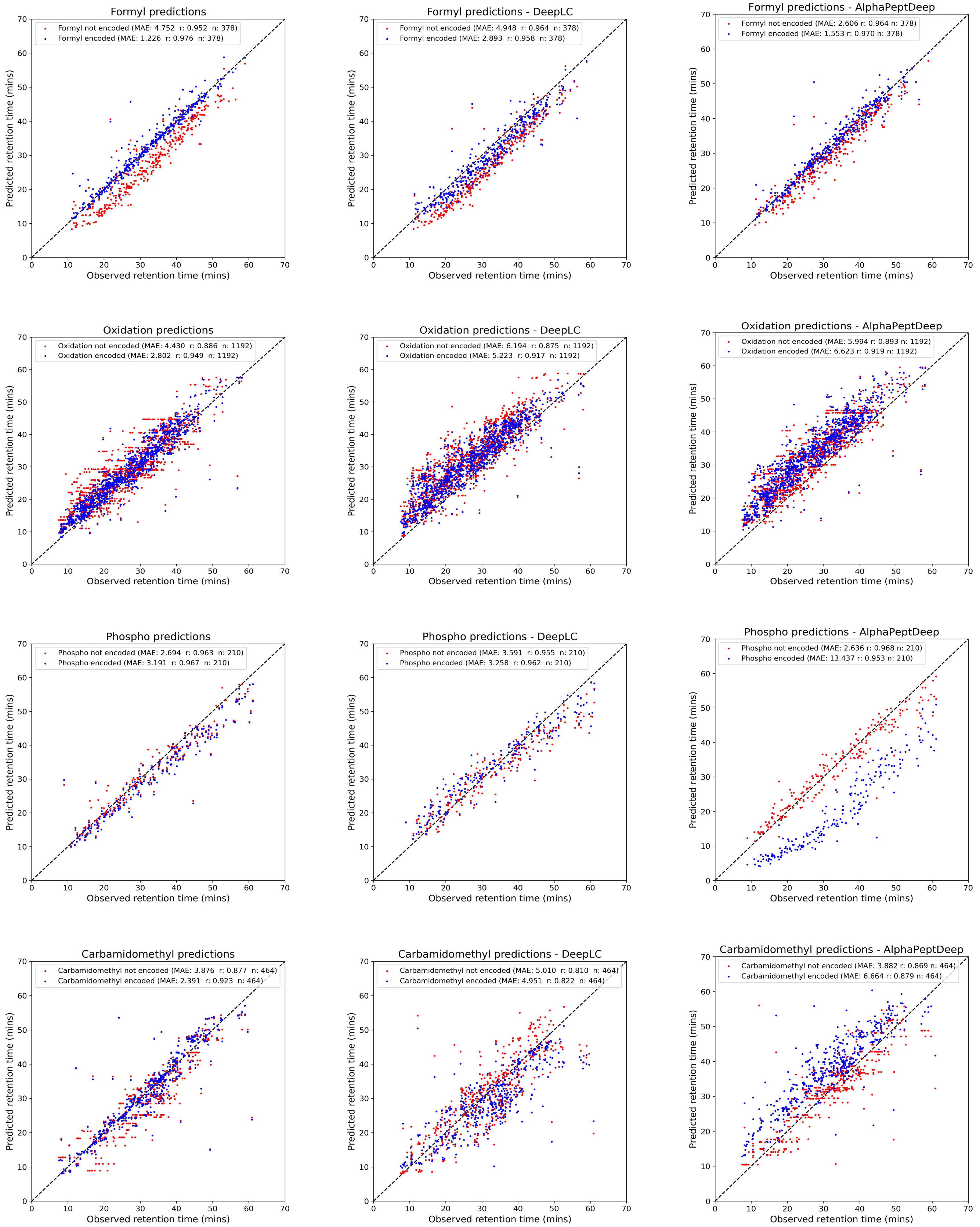

**Supplementary Figure 5.** Plot of predicted vs. true RT for the modifications formyl, oxidation, phospho and carbamidomethyl in each row, using GraphRT, DeepLC and AlphaPeptDeep in each column.

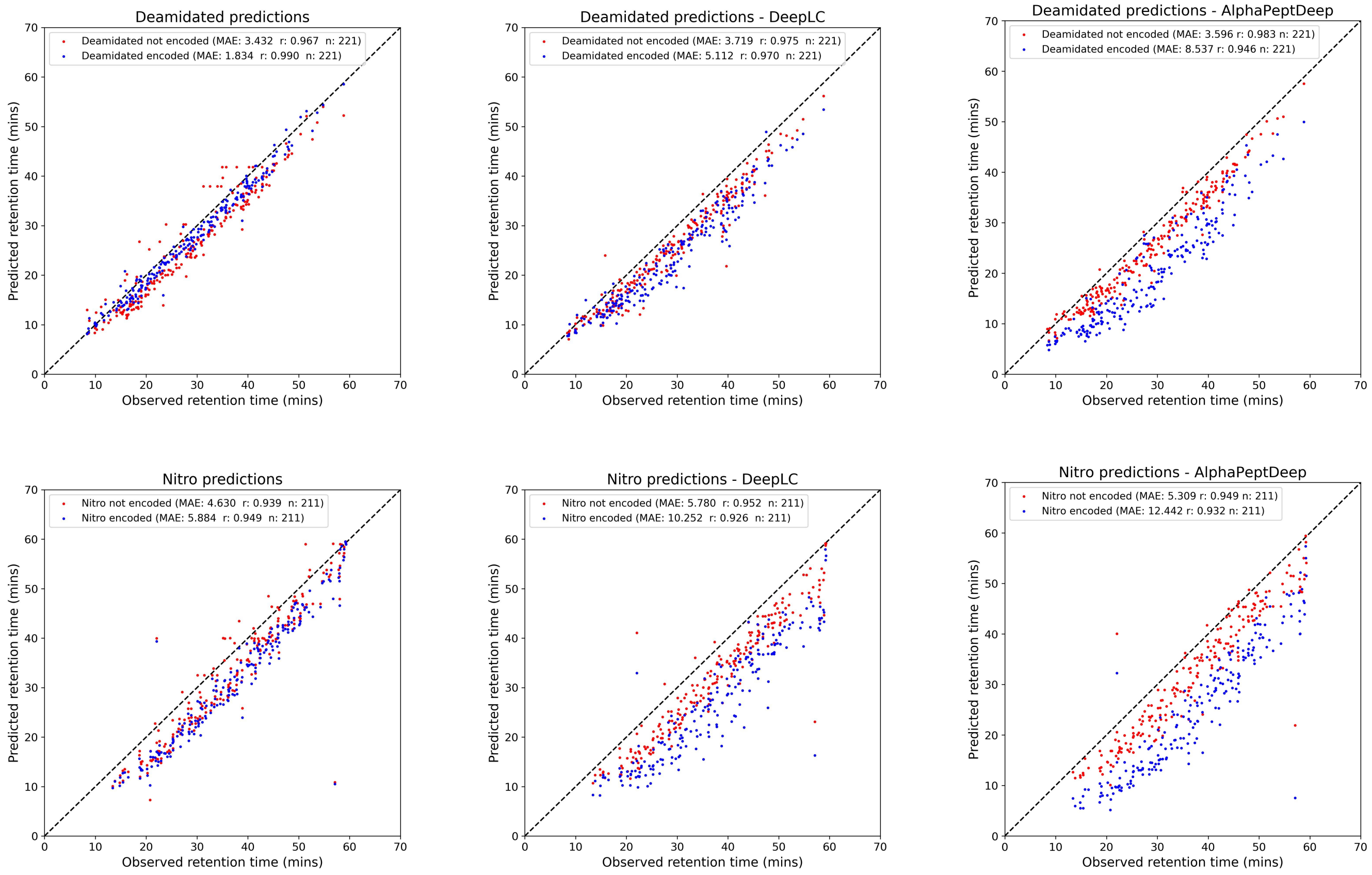

**Supplementary Figure 6.** Plot of predicted vs. true RT for the modifications deamidated and nitro in each row, using GraphRT, DeepLC and AlphaPeptDeep in each column.

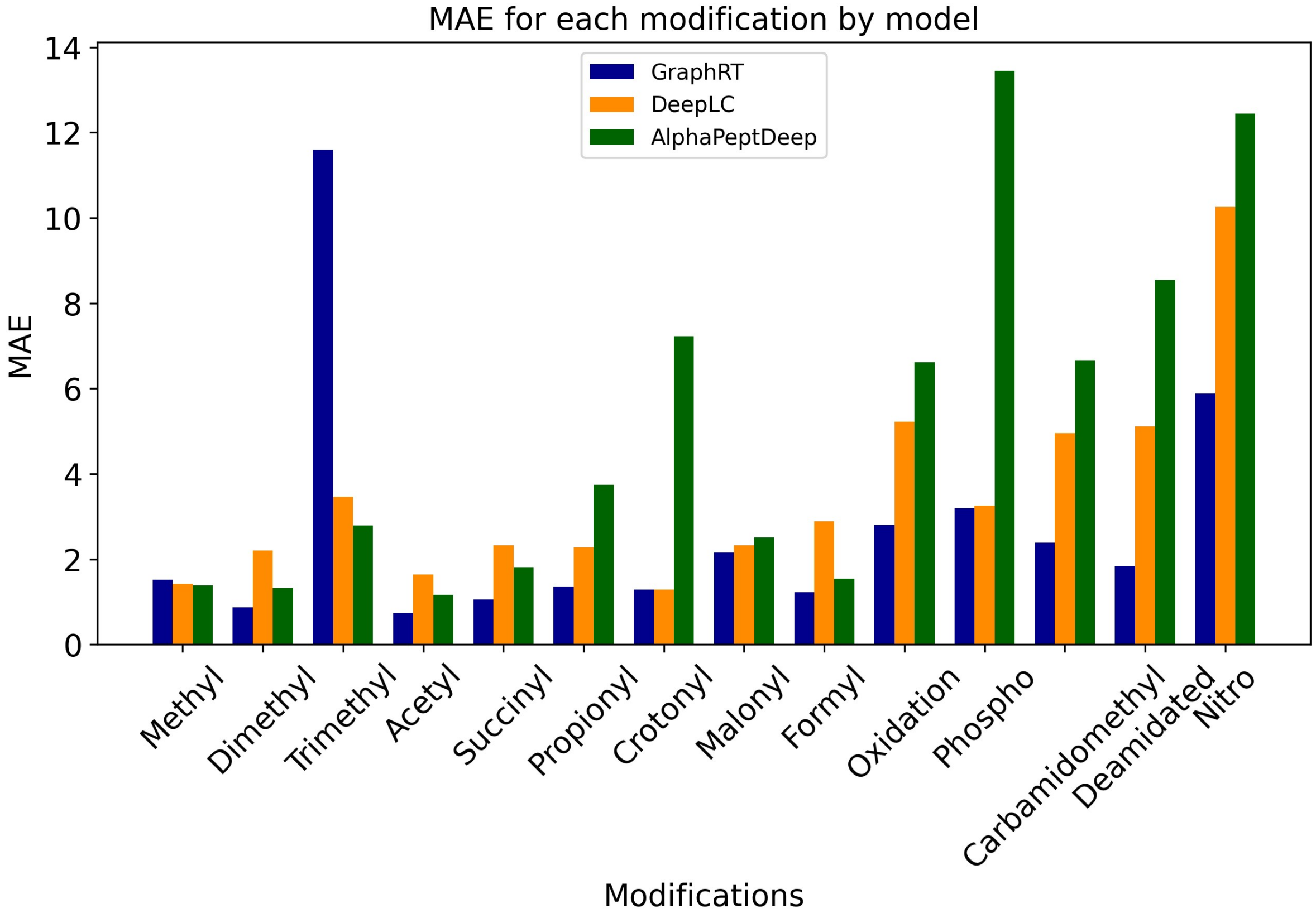

**Supplementary Figure 7.** Bar chart summarising the table and plots above by showing all 14 modifications and the corresponding MAE values in the encoded test set for GraphRT (dark blue) and DeepLC (orange). GraphRT outperforms DeepLC in terms of MAE for 12 of the 14 modifications.

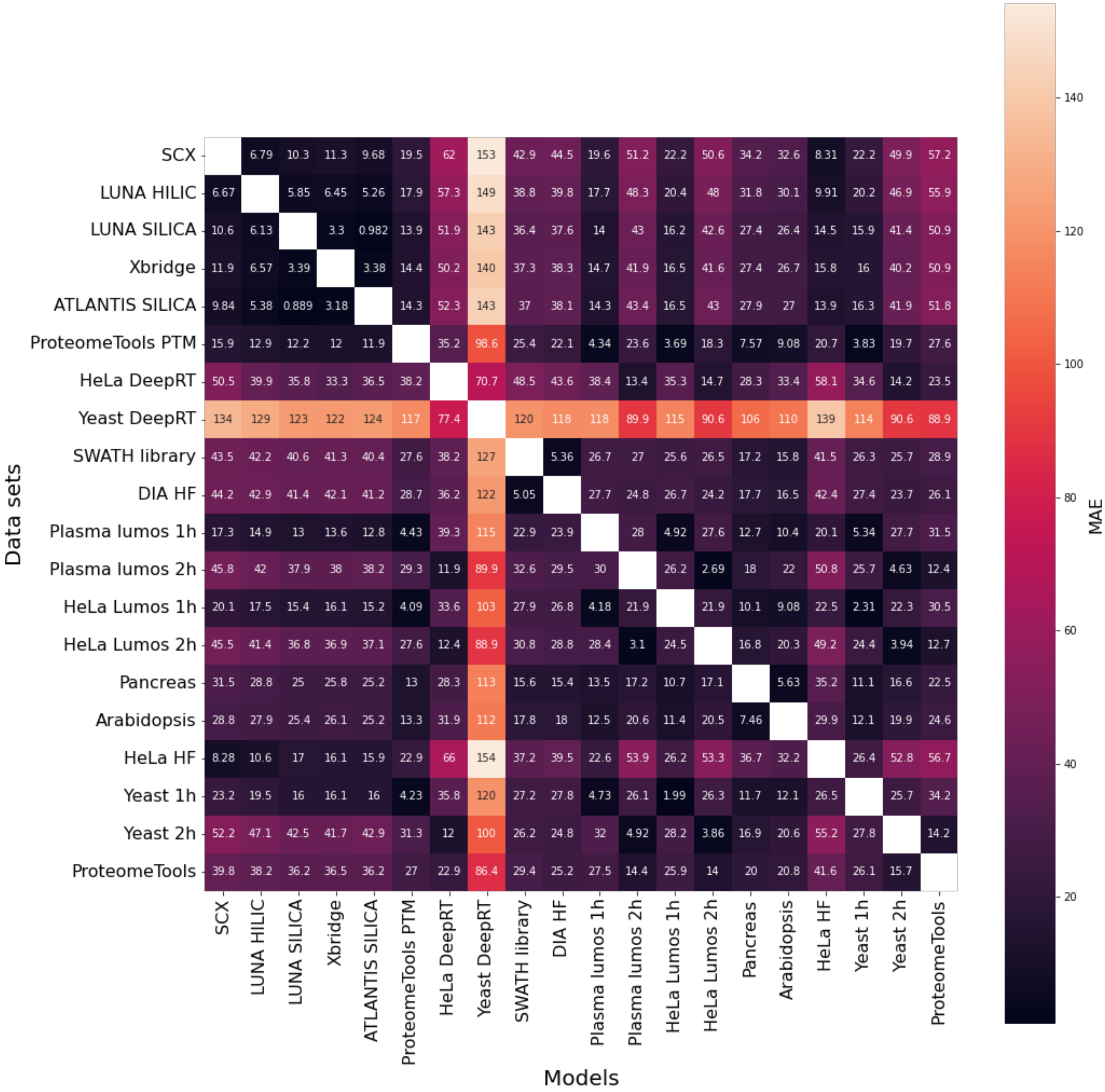

**Supplementary Figure 8**. Heatmap showing the MAE of each version of GraphRT pre-trained on a particular data set, tested on one of the 19 other data sets.

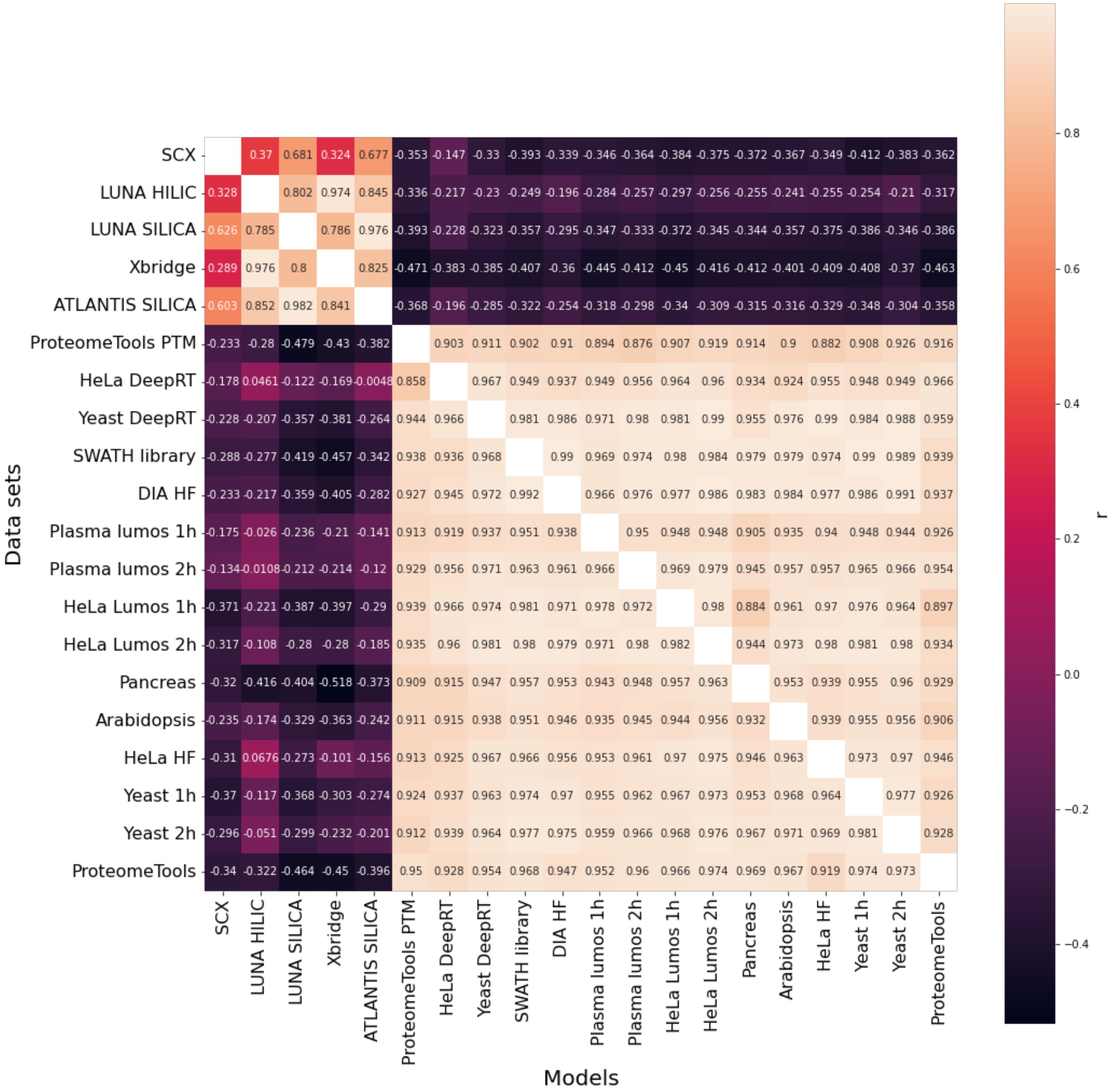

**Supplementary Figure 9**. Heatmap showing the r of each version of GraphRT pre-trained on a particular data set, tested on one of the 19 other data sets.